\documentclass[conference]{IEEEtran}
\IEEEoverridecommandlockouts
\def\IEEEtitletopspace{18pt}
\usepackage{cite}
\usepackage{amsmath,amssymb,amsfonts}
\usepackage{algorithmic}
\usepackage{graphicx}
\usepackage{textcomp}
\usepackage{xcolor}
\usepackage{comment}
\usepackage{tikz}
\usepackage{aeguill}
\usepackage[frozencache,cachedir=.]{minted}
\usepackage{enumitem}
\usepackage{parskip}
\def\BibTeX{{\rm B\kern-.05em{\sc i\kern-.025em b}\kern-.08em
    T\kern-.1667em\lower.7ex\hbox{E}\kern-.125emX}}
\begin{document}

\title{Formalizing Operational Design Domains\\ with the Pkl Language
\thanks{We acknowledge the support of the Swedish Knowledge Foundation via the industrial doctoral school RELIANT, grant nr: 20220130. This research was carried out within the SUNRISE project and is funded by the European Union's Horizon Europe Research and Innovation Actions under grant agreement No.101069573. However, views and opinions expressed are those of the author(s) only and do not necessarily reflect those of the European Union or the European Union's Horizon Europe Research and Innovation Actions.}
}
\makeatletter
\newcommand{\linebreakand}{%
  \end{@IEEEauthorhalign}
  \hfill\mbox{}\par
  \mbox{}\hfill\begin{@IEEEauthorhalign}
}
\makeatother

\author{
\IEEEauthorblockN{Martin Skoglund}
\IEEEauthorblockA{\textit{RISE Research Institutes of Sweden} \\
Borås, Sweden \\
martin.skoglund@ri.se}
\and
\IEEEauthorblockN{Fredrik Warg}
\IEEEauthorblockA{\textit{RISE Research Institutes of Sweden} \\
Borås, Sweden \\
fredrik.warg@ri.se}
\and
\IEEEauthorblockN{Anders Thorsén}
\IEEEauthorblockA{\textit{RISE Research Institutes of Sweden} \\
Borås, Sweden \\
anders.thorsen@ri.se}
\linebreakand
\IEEEauthorblockN{Hans Hansson}
\IEEEauthorblockA{
\textit{MRTC, Mälardalen University}\\
Västerås, Sweden \\
hans.hansson@mdu.se}
\and
\IEEEauthorblockN{Sasikumar Punnekkat}
\IEEEauthorblockA{
\textit{MRTC, Mälardalen University}\\
Västerås, Sweden \\
sasikumar.punnekkat@mdu.se}
}

\maketitle

\begin{abstract}  
The deployment of automated functions that can operate without direct human supervision has changed safety evaluation in domains seeking higher levels of automation. Unlike conventional systems that rely on human operators, these functions require new assessment frameworks to demonstrate that they do not introduce unacceptable risks under real-world conditions.
To make a convincing safety claim, the developer must present a thorough justification argument, supported by evidence, that a function is free from unreasonable risk when operated in its intended context. The key concept relevant to the presented work is the intended context, often captured by an Operational Design Domain specification (ODD).
ODD formalization is challenging due to the need to maintain flexibility in adopting diverse specification formats while preserving consistency and traceability and integrating seamlessly into the development, validation, and assessment. 
This paper presents a way to formalize an ODD in the Pkl language, addressing central challenges in specifying ODDs while improving usability through specialized configuration language features. The approach is illustrated with an automotive example but can be broadly applied to ensure rigorous assessments of operational contexts.
\end{abstract}


\section{Introduction} \label{sec:intro}
Ensuring the safety of automated functions remains a key challenge for widespread deployment, as these systems must reliably manage a broad spectrum of operating conditions. The complete set of situations an automated function must handle safely can be framed by all dynamic interactions in the operating context, i.e., all permutations over all conceivable interaction scenarios in the complete intended operational design domain (ODD). This often involves truncating the long tail of increasingly improbable situations where the likelihood of omitted hazardous situations posing an unreasonable risk is sufficiently low.
Building on scenario-based testing practices specific to the automotive domain from which the example ODD taxonomy originates, ISO 34501~\cite{iso34501} defines the ODD as "the operating conditions under which a given driving automation system or feature thereof is specifically designed to function, including but not limited to, environmental, geographical, and time-of-day restrictions and/or the requisite presence or absence of specific traffic or roadway characteristics." 

Validating the vast set of scenarios in context becomes increasingly complex at higher levels of automation, requiring exploration of growing test spaces to provide evidence to substantiate the safety claims. In the automotive field, this is commonly called the "billion-miles" challenge~\cite{kalraDrivingSafetyHow2016a} but extends to any domain with automation ambitions.

An ODD isolates a subset of driving conditions under which an automated function must have safe operations. A formally defined ODD helps streamline test coverage analysis, validation, and assessment by specifying a subset of driving conditions within which the system's safe operation must be assured. By clearly specifying the boundaries and conditions under which an automated function operates, the ODD helps to narrow the scenario space to the essential contexts. Rather than attempting to account for an infinite range of real-world conditions, we establish a well-defined set of parameters that can be systematically evaluated and refined for both development teams and safety assessors. As a result, the total physical testing needed can be substantially reduced—moving towards a solution to the scale of the "billion-mile" problem.

Although existing taxonomies and standards (e.g., ISO 34503 and DOT HS 812 623\cite{iso34503, NTHSA2018}) offer structured approaches to defining ODDs, they typically rely on natural language and do not provide a standardized exchange mechanism. These limitations make it challenging to integrate them into automated testing toolchains. While standardizing ODD semantics is essential, the work here focuses on creating an infrastructure that optimizes usability, drawing on objectives that appear in draft or conceptual work on standardizing ODDs and test toolchain integration, i.e., ASAM OpenODD~\cite{OpenODD} and ASAM Open Test Specification~\cite{OpenTestSpec}.

This work does not seek to develop a new ODD taxonomy. Instead, it introduces a formalization and a configuration-based mechanism for specifying and exchanging ODDs. The presented approach is illustrated using ISO 34503 as taxonomy and automotive applications as representative use cases without constraining the method to those particular examples. The formalization leverages the Pkl language~\cite{Pkl} to address the core challenges of ODD specification, focusing on flexibility, extensibility, and traceability. By integrating these features, the presented contribution facilitates creating and managing ODD definitions and ensures that tests can be readily linked to operational conditions throughout development, validation, and assessment. 


\section{Background and Related Work} \label{sec:background}

Standardizing ODD definitions is essential for safety assurance in automated driving systems. Several standards, such as ISO 34503~\cite{iso34503} and BSI PAS 1883~\cite{PAS18832020a}, as well as ongoing projects like ASAM OpenODD~\cite{OpenODD} and OpenTestSpecification~\cite{OpenTestSpec}, respond to this need by providing structured approaches to describe ODDs. Among these efforts, interoperability remains a critical objective. However, these undertakings are evolving, but there is still limited guidance on fully integrating an ODD definition process with robust test workflows or real-time monitoring~\cite{charmet_monitor}. 

Prior studies have explored various strategies for defining and applying ODDs across different contexts. Lee et al.~\cite{leeIdentifyingOperationalDesign2020a} propose a statistical approach that identifies an ODD by mapping geographic regions with acceptable risk levels, enabling a data-driven method of boundary delineation. Erz et al.~\cite{erzOntologyThatReconciles2022a} focus on ontology-based techniques to reconcile ODD definitions with vehicle architectures and scenario-based testing, providing structured cross-referencing throughout system development. Sun et al.~\cite{sunAcclimatizingOperationalDesign2022a} introduce formal specification methods to manage ODD extraction during automated driving system development. Further work looks at the role of ODD metrics in evaluating system maturity and usability~\cite{kaiserDefinitionMetricsAssessment2023a}, as well as risk-based scenario coverage~\cite{weissensteinerODD2023}, illustrating the breadth of ongoing efforts to refine ODDs for comprehensive safety assurance. The relevant literature primarily addresses the content of ODDs or interactions with the surrounding environment rather than a generalized formalization. To verify the practicality of our approach and ensure that it builds upon the existing body of work, we aim to define a set of evaluation criteria for our proposed solution. 

Pkl, pronounced "Pickle," is a configuration-as-code language developed by Apple, which provides robust validation and tooling capabilities~\cite{Pkl}. It can be used as a command-line tool, a software library, or a build plug-in, making it adaptable to a wide range of use cases. It is designed to scale smoothly from small, simple, and ad-hoc tasks to large, complex, and recurring configuration challenges. In this work's context, Pkl aims to ensure consistent configurations across multiple development environments, emphasizing automation and integration into DevOps pipelines. It was developed to handle the product line configuration of Apple smartphones. 

At its core, Pkl implements a special-purpose configuration language that combines the benefits of a static configuration format with the flexibility of a general-purpose programming language. This hybrid approach is well-suited for generating static configurations, such as those formatted in JSON, YAML, or other structured data formats. By leveraging Pkl, users can automate the creation, validation, and maintenance of configurations for tools and systems, ensuring consistency and scalability across diverse applications.

In the present work, ISO 34503~\cite{iso34503} serves as the basis for defining the ODD, aligning with international standards and offering a consistent way to specify ODD attributes and boundaries in the automotive domain. It is the specification stage conditions under which the system is intended to operate are specified. 
Several other practices have been identified to have an impact on how to treat the formalization of the ODD; ISO 29119~\cite{iso29119_2022} provides a comprehensive software and system testing framework, covering all aspects of the testing process and offering clear guidelines for designing, executing, and documenting tests. The functional safety standard ISO 26262~\cite{iso26262} addresses functional safety in the automotive domain by guiding hazard analysis and risk assessment to the intended context of ODD. Safety of the Intended Functionality (SOTIF) ISO 21448~\cite{SOTIF}  extends safety considerations beyond functional safety, accounting for functional insufficiencies in implementing real-world operating conditions. Together, the practices inform the formalization design of ODD interactions that should be considered in test case allocation, test environment setup, and test data requirements, bridging the conceptual design of the ODD with practical testing and assessment applications.

We extend the ODD definition through a formalization using the Pkl framework. Formalization here means representing the ODD's attributes, constraints, and relationships in a machine-readable manner that supports automated querying and validation and enables integration into existing development toolchains. 
Our goal is flexibility and traceability in specifying and modifying ODDs. The approach aims to facilitate direct linkage between ODD descriptions and verification or assessment processes, reducing ambiguity while supporting systematic evidence gathering for safety assurance~\cite{gyllenhammarODD2020}. 

\section {Evaluation criteria for an ODD formalization} \label{sec:spec}
The following criteria are defined based on insights obtained from Section \ref{sec:background} and further enriched by discussions with interaction experts in the assessment field, ensuring a solid foundation for analysis. They serve as a basis for judging whether the attempt at formalization is valuable. The evaluation criteria are semi-domain-specific but can be reformulated to become relevant to any domain. For example, Tonk et al.~\cite{tonkSpecifiedOperationalDesign2021b} demonstrated that ODD principles can be adapted for remote driving in the railway domain, thus illustrating how the concept extends beyond automotive use cases.

The following core criteria were identified to enable a flexible and efficient use of ODDs:
\begin{enumerate} [label=\textbf{CC\arabic*}]
        \item Abstraction of the complex structure to enable treatment of specific aspects under consideration. \label{c1}
        \item Templating reduces duplication and maintains internal consistency, thus aiding flexibility and reducing mistakes while accommodating probabilistic qualifiers. \label{c2}
        \item Validation can parse and check validity automatically. The solution should incorporate conditional statements to precisely express constraints or dependencies and support inclusive classification for each condition. \label{c3}
\end{enumerate}

Safety artifacts like the ODD  directly influence functional behavior and the criteria used to evaluate that behavior, necessitating safeguards that ensure integrity, traceability, and control safeguards. For example, version control should document all modifications, along with their rationale and approvals. Validation reviews must ensure that updates meet safety criteria and do not create additional risks, while impact analyses assess how changes affect system safety and compliance. These measures help maintain the reliability of safety-critical artifacts and align them with relevant safety standards.

The following safety criteria related to artifacts according to ISO 26262-8:2018~\cite{iso26262}:
    \begin{enumerate} [label=\textbf{SC\arabic*}]
        \item Integrity provides measures that keep the ODDs corruption-free, like unintended modification of attributes or constraints. \label{c4}
        \item Human readability to enable confirmation reviews. ODD parameters must be expressed so relevant stakeholders can understand, allowing for unambiguous review, approval, and justification. This aligns with ISO 26262's emphasis on comprehensibility, unambiguity, and ease of maintenance. \label{c5}
        \item Specification and Management of Safety Requirements. Following ISO 26262-8:2018, an ODD language must support proper safety requirement attributes: unique identification, status tracking, criticality level, clear scope, and allocated ownership. This includes mechanisms for traceability back to source requirements and forward to derived or lower-level elements, ensuring internal and external consistency, completeness, and maintainability throughout the lifecycle. \label{c6}
    \end{enumerate} 

Input criteria for the ODD are that it must accommodate adaptability to ensure maintainability. This includes supporting a flexible architecture for communal refinement and alignment with evolving ODD taxonomies. Maintainability also requires that updates—whether to the ODD itself or related taxonomies—can be efficiently tracked, reviewed, and implemented.
\begin{figure}
    \centering
    \includegraphics[width=0.5\textwidth]{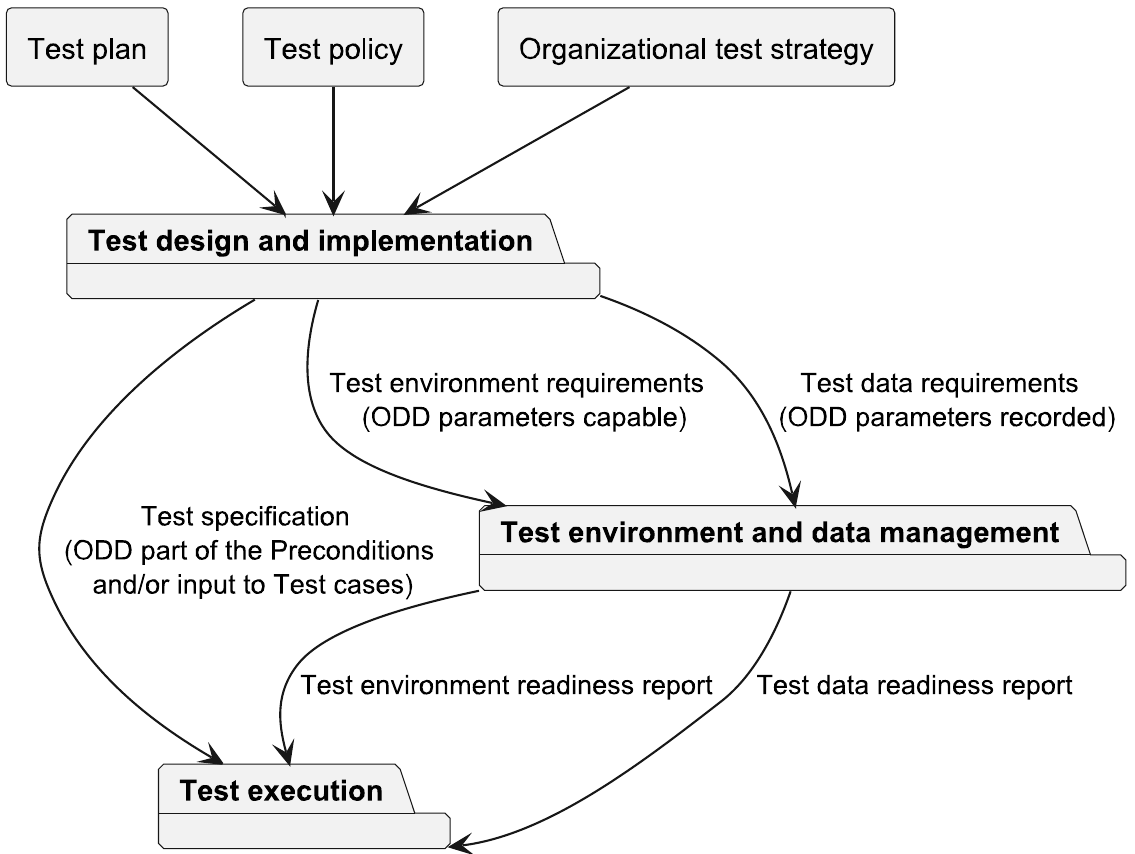}
    \caption{Dynamic Test Process ISO 29119:2022, how the ODD can integrates with the overall testing process~\cite{iso29119_2022}.}
    \label{fig:test}
\end{figure}
This ensures that modifications remain transparent and traceable, enabling controlled evolution without compromising consistency or safety assurance. The output criteria for the ODD include seamless integration with general dynamic testing frameworks, as outlined in standards such as ISO 29119~\cite{iso29119_2022}. In Fig. \ref{fig:test}, the test specification encompasses the complete documentation of the test design, test cases, which contain a concrete ODD, and test procedures for a specific test item. Suppose such details are not included in the test plan. Any additional requirements for the test environment, which contains ODD capabilities and test data partly connected to ODD, must also be specified.
This permissive definition generally involves more specific scenario-based testing approaches, such as those in ISO 34502~\cite{iso34502}, and simulation-oriented frameworks, such as those under development in the ASAM Test Specification project. These requirements demonstrate the broader applicability of the ODD formalization to similar contexts or domains that rely on structured, testable operational boundaries.
It can be seen as input and output capabilities grouped as interface criteria in the formulation of the ODD:
    \begin{enumerate} [label=\textbf{IC\arabic*}]
        \item Taxonomy Integration - The language must support the integration of evolving or domain-specific taxonomies, ensuring compatibility and consistency in the definitions of operational constraints. \label{c7}
        \item Test Execution Enrichment - ODD descriptions should aid in scenario-based validation by enabling dynamic checks of whether a scenario remains within bounds and by supporting the generation of concrete test cases aligned with the ODD constraints. \label{c8}
    \end{enumerate}

On a side note, the permissive ODD definition described, for instance, in the OpenODD concept paper~\cite{OpenODD} will not be considered at this time. A permissive definition generally includes everything within the ODD except for explicitly excluded elements, whereas a restrictive definition includes only those explicitly specified elements.

A permissive ODD employs a define-by-exclusion approach yet does not compute the set difference from a universal set; instead, it works directly with the ODD set, resulting in an inverse form of a restrictive ODD. Removing specific elements from an otherwise complete set can be replicated using default values and templating methods, which characterize similar configuration elements by focusing on their differences. Although there is no plan to adopt a strictly permissive definition, the implementation could be adjusted to accommodate that approach.

\section{ODD Formalization} \label{sec:odd}
In this section, we describe the hierarchical taxonomy in the ISO 34503 ODD definition, illustrated in Fig. 2, using the features of a modern configuration language, Pkl [7], although our ODD is restrictive since parameters must be explicitly specified.

\begin{figure}
    \centering
    \includegraphics[width=0.5\textwidth]{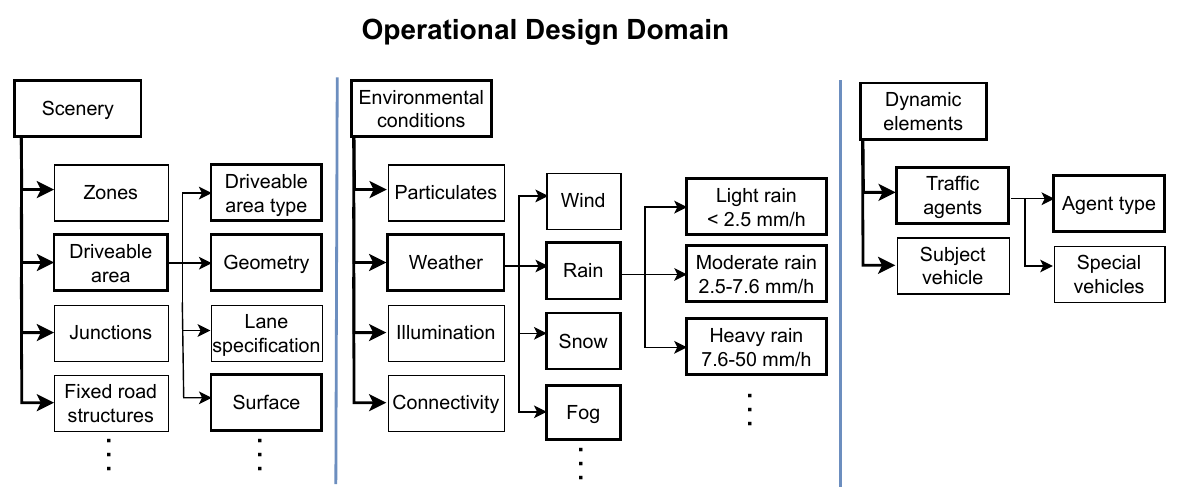}  
    \caption{Part of the ODD Taxonomy from ISO 34503~~\cite{iso34503}.}
    \label{fig:oddtaxonomy}
\end{figure}

Pkl allows modules to be split into smaller, more manageable parts through imports, which can come from local or remote locations. The hierarchy of artifacts may look like this: ODD-template -> Instance of ODD from template -> Instance of one configuration of ODD elements. A configurable security policy helps control these imports, allowing complex configurations to be refined or extended without compromising integrity. The constructed ODD templates in Pkl can be found in~\cite{gitODD}. The top module is shown in Fig. \ref{alg:oddtop}.

\begin{figure}

\begin{minted}[xleftmargin=2em, mathescape, linenos, fontsize=\small]{python}
# Note: Template for ISO34503 ODD 

@ModuleInfo { minPklVersion = "0.25.1" }
module ODD.ODD_template.pkl

import "dyn_template.pkl"
import "env_template.pkl"
import "scen_template.pkl"

class odd {
  scenery : scen_template.scenery
  environment: env_template.env
  dynamic : dyn_template.dynamic_elements
}
\end{minted}
\caption{Top module for ISO 34503 ODD in Pkl.}\label{alg:oddtop}
\end{figure}

A template is natively reusable; all hierarchies are preserved, where objects and entire modules can be defined partially and then turned into concrete configurations following a schema and by specifying or overriding defaults. Sharing these templates among teams or across networks streamlines the creation of consistent, reusable configurations. The version of the evaluation toolchain can be stipulated in the template module, forcing alignment with the feature and the configuration data. This approach reduces duplication, enhances maintainability, and aligns well with scenarios that require frequent updates or collaboration among multiple stakeholders. 

\begin{figure}
\begin{minted}[xleftmargin=2em, mathescape, linenos, fontsize=\small]{python}
# Note: Excerpt from the file scen_template.pkl 

const speed_limit_global = 30.0

typealias Direction_of_travel = "right_hand_travel"
| "left_hand_travel"

class Lane_dimensions {
  // Define properties related to lane dimensions
  lane_dimension : Float (isBetween(2.7, 3.2)) = 2.7 
  // meters
}

class Drivable_area_lane_specification {
  lane_dimensions: Lane_dimensions
  lane_markings: Lane_markings
  lane_type: Lane_type
  direction_of_travel: Direction_of_travel
  speed_limit : Float (isBetween(0, 
  speed_limit_global))  
  lane_usage : Boolean = true                             

}
\end{minted}
\caption{Example: Definition of drivable area.}\label{alg:driveablearea}
\end{figure}

In the definition of the ODD attribute \textit{Drivable Area}, as shown in Fig.~\ref{alg:driveablearea}, typed objects are derived from templates, permitting only amendment by override of existing properties rather than the creation of new ones. This setup allows the specification of lane dimensions to be amended based on different \textit{drivable\_area\_type}. For instance, if trucks are present in the ODD, lane width must exceed 2.6 meters~\cite{FederalSizeRegulationsa}, and the dimension is enforced to fall within a valid range with a suitable default.
While not explored in this base configuration, Pkl can handle arbitrary elements, and there are several ways to group elements and values, for example, as Listings, which is an ordered, indexed collection of elements, or as values in Mappings, which is an ordered collection indexed by a key. For example, a Listing could express probabilities, e.g. (name = "event1"; type = "failure"; probability = 0.6). 

\begin{figure}
\begin{minted}[xleftmargin=2em, mathescape, linenos, fontsize=\small]{python}
# Note: From the file ODD1_test.pkl importing
# and configuring ODD_template.pkl 

import   "ODD_template.pkl"
 odd1 : ODD_template.odd =  new {
   scenery {
     zone {
       region_or_state = "Sweden"
     }
     drivable_area {
       drivable_area_lane_specification {
         direction_of_travel = "right_hand_travel"
         speed_limit = 15.0 
         lane_usage = true
       }
     }
   }
 }
\end{minted}
\caption{Instantiating an ODD from the template.}\label{alg:oddconfiguration}
\end{figure}

Only things lacking default values must be configured when spawning a complex ODD. It is desirable to explicitly configure parameters of particular interest or critical safety relevance for the ODD. In the example shown in Fig. \ref{alg:driveablearea}, we can see that direction\_of\_travel is an enumerated class with no default value. So, when constructing an ODD from this template, the direction of travel must be configured as either right\_hand\_travel or left\_hand\_travel, as shown in the example instantiation of \textit{odd1} in Fig. \ref{alg:oddconfiguration}. 

\begin{figure}
\begin{minted}[xleftmargin=2em, mathescape, linenos, fontsize=\small]{python}
#Note Pkl can render JSON, Jsonnet, 
#Pcf (a static subset of Pkl), (Java) Properties,
#Property List, XML, YAML

C:\pkl> ./pkl eval -f json .\ODD1_test.pkl   
...
    "drivable_area_lane_specification": {
      "lane_dimensions": {
        "lane_dimension": 2.8
      },
      "lane_markings": {
        "clear_lane_marking": true,
        "blurred_lane_marking": false,
        "no_lane_marking": false,
        "temporary_lane_marking": false
      },
      "lane_type": {
        "bus_lane": false,
        "traffic_lane": true,
        "cyclists_lane": false,
        "tram_lane": false,
        "emergency_lane": false,
        "shared_lane": false,
        "other_special_purpose_lane": false
      },
      "direction_of_travel": "right_hand_travel",
      "speed_limit": 15.0,
      "lane_usage": true
    },
...
\end{minted}
\caption{Rendering an ODD as JSON.}\label{alg:oddjson}
\end{figure}

\begin{figure} [h!]
    \centering
    \includegraphics[width=0.45\textwidth]{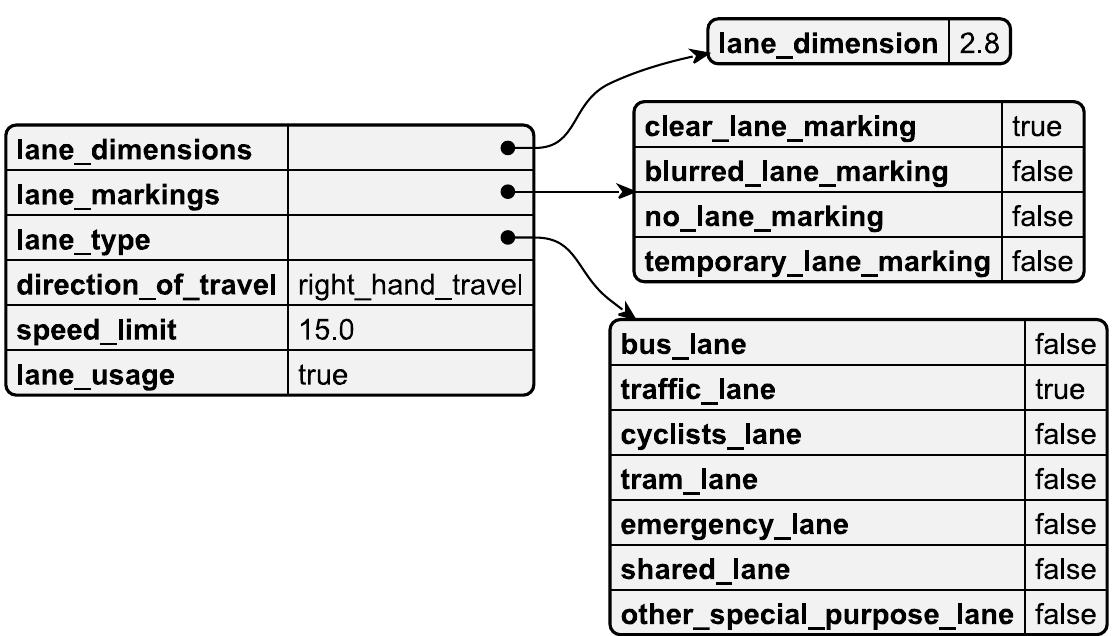}
    \caption{The rendered part of the JSON ODD, excerpt of Fig. \ref{fig:complete}, visualized by PlantUML~\cite{plantuml}. }
    \label{fig:tree}
\end{figure}

Data in Pkl is not manipulated directly; when a value is modified, it is a new value, leaving the original value intact. This immutability feature helps reduce errors. The evaluation of Pkl code is strictly confined to a sandbox environment, with no external interactions except for narrowly defined and controlled cases. The framework can be initialized with automated defaults, thereby simplifying setup. Additionally, validation enforces strict domain-specific constraints to ensure that each parameter remains consistent with the operational requirements of the defined ODD. 
A more assessable representation of the data can be obtained by exporting the configuration into a static format such as JSON or YAML and converting it into a tree structure diagram with the markup framework PlantUML~\cite{plantuml}. Given a hierarchical visualization's clarity, this approach results in a graphical format that is easier to review and interpret. Fig. \ref{alg:oddjson} shows how an ODD is rendered in JSON; Fig. \ref{fig:tree} (again an excerpt of the drivable area) and Fig. \ref{fig:complete} (complete ISO 34503 ODD) shows graphical representations rendered by PlantUML. The intention is not to present all the details but to convey a sense of the complete structure. 

\begin{figure} [h]
\begin{minted}[xleftmargin=2em, mathescape, linenos, fontsize=\small]{python}
# Note: speed_limit_global = 30.0

C:\pkl> ./pkl eval .\ODD1_test.pkl

Type constraint 'isBetween(0, speed_limit_global)' 
violated.
Value: 31.0

139 | speed_limit :
Float (isBetween(0, speed_limit_global)) 
      ^^^^^^^^^^^^^^^^^^^^^^^^^^^^^^^^
at ODD.scen_template.pkl
#Drivable_area_lane_specification.speed_limit 
(file:///C:/pkl/scen_template.pkl, line 139)
\end{minted}
\caption{Trying to generate an ODD with lane speed 31, which is outside limits.}\label{alg:outsidelimits}
\end{figure}

If all type constraints are satisfied, the Pkl evaluator converts the data model into an external representation and terminates with a zero status code; otherwise, it outputs an error message. Fig. \ref{alg:outsidelimits} shows an example where an instantiated ODD uses a speed limit outside the valid range defined in the template. 

\section{Evaluation}
Reducing duplication and maintaining consistency is necessary to formalize an ODD flexibly and efficiently. The following points illustrate how the approach addresses these requirements:
\begin{itemize} 
\item \ref{c1} is fulfilled by applying abstractions. The efforts can be concentrated on important aspects of the ODD; for example, all parameters that are not in focus can be given default values. 
\item \ref{c2} is fulfilled, subdividing large ODDs into specialized modules (for example, scenery, environment, and dynamic objects) enhances clarity and assists distributed teams in maintaining correctness, thus minimizing errors arising from rewriting similar elements.
\item \ref{c3} is fulfilled, robust validation mechanisms enable automatic checks of configuration attributes using conditional statements, numeric ranges, and enumerated types.
\end{itemize}

Additional design considerations extend the approach to feature important safety artifacts:
\begin{itemize} 
\item \ref{c4} is fulfilled by immutability and isolation to ensure unintended modifications to ODD objects cannot happen. This design safeguards integrity. 
\item \ref{c5} is partially fulfilled as the readability is satisfied through the possibility of rendering the resulting configuration in a graphical tree structure; see Fig. \ref{fig:tree}. In addition, Pkl code is deliberately designed to match the configuration it produces.
The formalization does not reference the taxonomy directly, and a delicate translation step is required to maintain the traceability and semantic meaning of parameters. 
\item \ref{c6} is partially fulfilled, though native traceability is lacking as Pkl and does not have built-in version management. Still, its structured data model integrates smoothly with external tools or version control systems. 
\end{itemize}

Finally, the approach accounts for external dependencies and testing aspects:
\begin{itemize} 
\item \ref{c7} is fulfilled by aligning ODD specifications with external ontologies and scenario-based testing. Pkl modules are deliberately designed to meet general configuration needs and can accommodate most ODD requirements by definition. 
\item \ref{c8} is partially addressed by providing a stable tool for ODD and real-time generation and validation. The language’s flexible structure allows for custom methods. Ongoing and future work will explore these options further, for example, by introducing mechanisms for comparing ODD instances.
\end{itemize}

\section{Conclusions} \label{sec:conclusions}

The work presented here demonstrates how an ODD can be formalized using the Pkl configuration language in a machine-interpretable format. The technical difficulty lies not necessarily in the basic Pkl syntax but in mapping and translating between the Pkl configuration and the conceptual ODD taxonomy to ensure alignment with system-level definitions. Additional challenges include integrating with external tools for version control and traceability, as required by safety standards, and developing the logic needed for dynamic runtime validation and integration with test scenario generation frameworks.
By evaluating key criteria like — abstraction, templating, validation, integrity, reviewable, traceability, adaptability, and versatility —this paper has shown that a Pkl-based formalization of ODDs can fulfill the identified criteria and serve as a versatile solution for different domains; an automated driving use case exemplifies this. Additionally, the Pkl language provides strong industrial and community backing, with a stable, well-maintained code base and rapid feature growth; albeit developed for other applications, the general configuration challenges remain the same. All things considered, adopting a Pkl-based approach for the formalization of ODDs presents a strong case. While a solution that meets all formalization criteria natively would offer an even stronger case, we are unaware of any such solution currently available.

We propose and make available~\cite{gitODD} an approach to representing the ODD~\cite{iso34503} in the Pkl configuration language. This representation may contain errors or missing elements. It is an initial proof of concept that may evolve through collective engagement. Its continued development relies on communal contributions and cooperative efforts, and it may gain broader acceptance if the research community deems it sufficiently valuable.     

Future work will investigate the feasibility of using the ODD to automate test scenario allocation, enabling the efficient distribution of test cases to the most appropriate virtual or physical environments. Additionally, the flexibility and extensibility of the proposed solution will be leveraged to construct an ODD tailored to the forestry domain. Efforts will also include assessing compatibility with the forthcoming OpenODD framework to ensure alignment with emerging ASAM standards to explore potential integration opportunities.



\bibliographystyle{IEEEtran}

\bibliography{./ref/SUNRISEInitalAllocation.bib}
\vspace{12pt}

\appendices  

\section{Complete ODD graph} \label{app2}

\begin{figure*}[tb]
    \begin{minipage}{0.95\textwidth}
        \centering
        \includegraphics[height=1.35\textwidth] {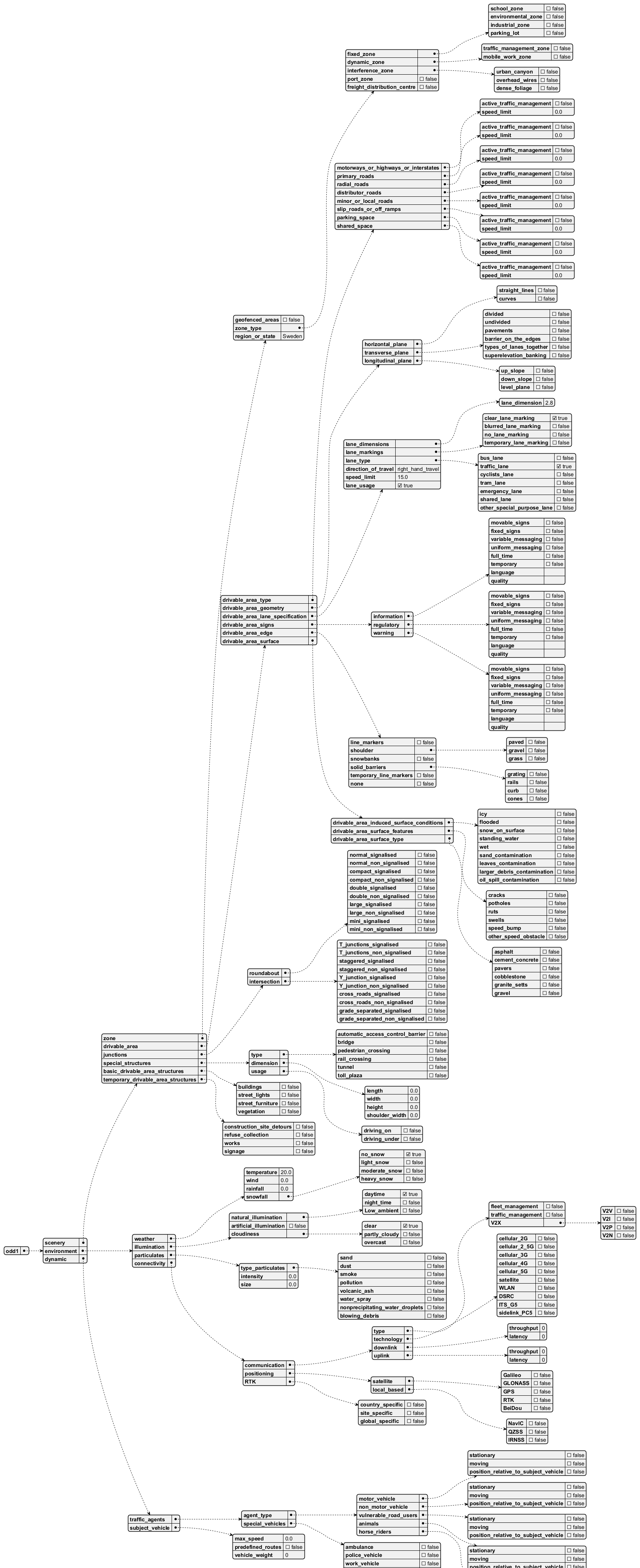} 
        \caption{Configured concrete ODD rendered in JSON visualized in PlantUML. Note: The intention is not to present all details but to convey a sense of the size and complexity of the complete structure.}
        \label{fig:complete}
    \end{minipage}\hfill
\end{figure*}

\end{document}